\documentclass[%
 reprint,
 amsmath,amssymb,
 aps,
]{revtex4-2}

\usepackage[pdftex]{graphicx}
\usepackage{dcolumn}
\usepackage{bm}

\begin{document}

\preprint{APS/123-QED}

\title{Exploration of a new reconstructed structure on GaN(0001) surface\\by Bayesian optimization}

\author{A. Kusaba}
 \email{kusaba@riam.kyushu-u.ac.jp}
\author{Y. Kangawa}
\affiliation{ 
Research Institute for Applied Mechanics, Kyushu University, Kasuga, Fukuoka 816-8580, Japan
}

\author{T. Kuboyama}
\affiliation{
Computer Centre, Gakushuin University, Toshima-ku, Tokyo 171-8588, Japan
}

\author{A. Oshiyama}
\affiliation{
Institute of Materials and Systems for Sustainability, Nagoya University, Chikusa-ku, Nagoya 464-8601, Japan
}

\date{\today}

\begin{abstract}
GaN(0001) surfaces with Ga- and H-adsorbates are fundamental stages for epitaxial growth of semiconductor thin films. 
We explore stable surface structures with nanometer scale by the density-functional calculations combined with Bayesian optimization, and succeed to reach a single structure with satisfactorily low mixing enthalpy among hundreds of thousand possible candidate structures. 
We find that the obtained structure is free from any postulated high symmetry previously introduced by human intuition, satisfies electron counting rule locally, and shows new adsorbate arrangement, reflecting characteristics of nitride semiconductors.
\end{abstract}

\maketitle


Nitride semiconductors which are already premier materials in optoelectronics \cite{amano1986metalorganic,nakamura1994candela} are now emerging in power electronics due to their superior physical properties to Si \cite{otake2008vertical}. 
GaN vertical metal-oxide-semiconductor devices show efficient energy conversion \cite{oka20151}, and \mbox{AlGaN/GaN} high-electron mobility transistors are also gradually spreading in the market \cite{palacios2005high}, inferring that our sustainable society will be realized with nitrides. 
However, to guarantee the expected high performance of such power devices, it is necessary to forge high-quality epitaxial films of GaN. 
Despite the many experimental efforts in this decade \cite{kachi2014recent,amano20182018}, the current quality of GaN crystalline films is unsatisfactory for replacing Si power devices in the market.

Improvement in the epitaxial growth technique has been pursued based on empirical knowledge and experimental trials and errors that require a huge amount of time and resources. 
It may be more promising to obtain knowledge of atomistic processes in growth phenomena first and then design optimum conditions for epitaxial growth. 
To unveil the growth phenomena, atom-scale identification of the surfaces on which atomic reactions take place is an essential prerequisite.
However, such identification is difficult in experiments since harsh environments, such as high temperatures during metal-organic vapor phase epitaxy (MOVPE), render sophisticated experimental techniques inapplicable. 
A computational approach based on the first principles of quantum theory is an alternative and reliable way to identify growth surfaces.

First-principles calculations within density-functional theory (DFT) have indeed been performed for the reconstruction of GaN surfaces \cite{rapcewicz1997theory,fritsch1998ab,zywietz1999adsorption,pignedoli2001dissociative,van2002first}. 
Furthermore, recognizing that the growth surface is in equilibrium with the gas phase with a particular temperature and partial pressure of each molecule under the growth condition, an approach combining DFT surface energies with thermodynamics of the gas phase has been proposed \cite{kangawa2001new}. 
By this approach, the phase diagram of the surface structures in the space of the growth temperature and the source-gas pressure is obtained, and our knowledge of the growth surface is expanded \cite{akiyama2012ab,kangawa2013surface,kempisty2019evolution}. 
Unfortunately, the conclusions obtained in the past are based on rather small-scale DFT calculations by imposing short-range periodicity, e.g., by using a (2$\times$2) lateral cell. 
What is really necessary is the identification of the atomic structures of a wide range, at least nanometer scale, of growth surfaces which presumably lack periodicities within the range.

\begin{figure}
\includegraphics[width=7.5cm]{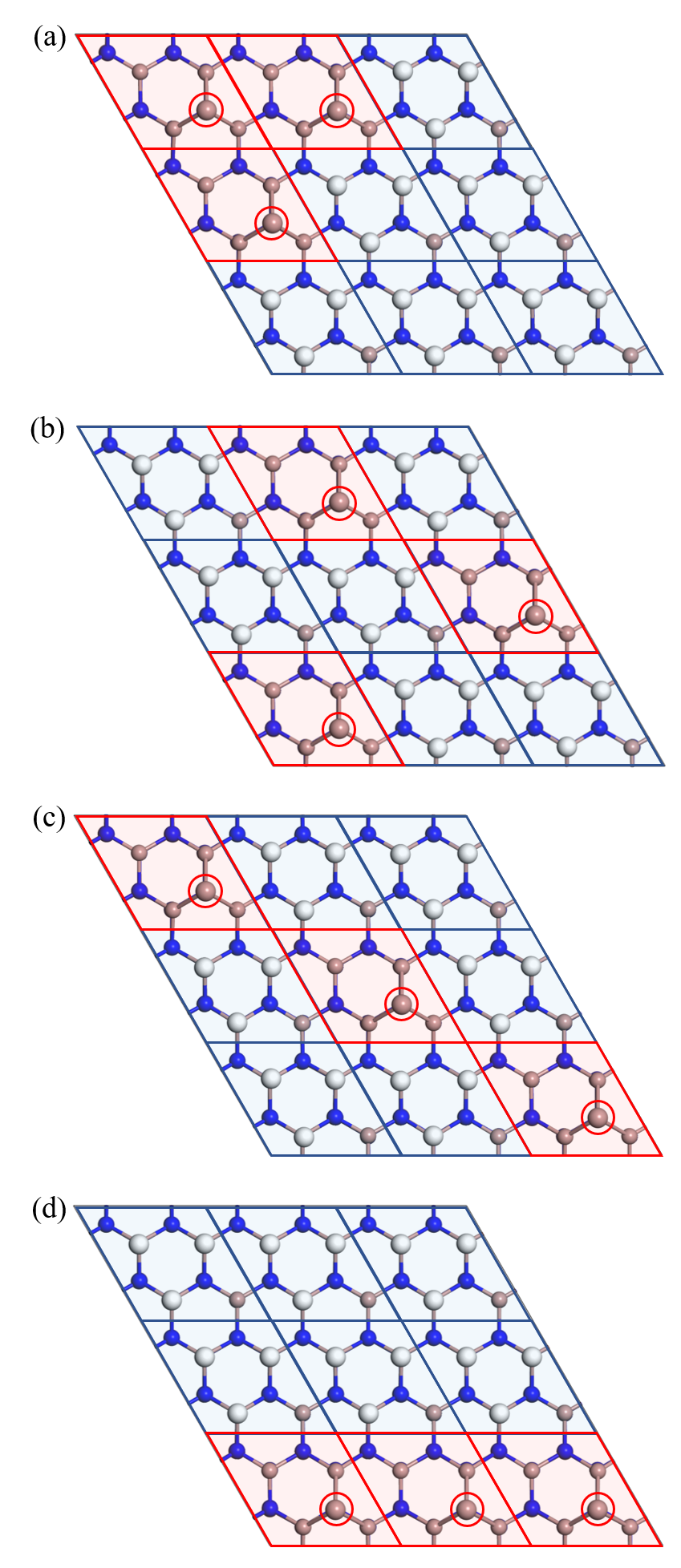}
\caption{\label{fig:baselines} Top views of the baseline (6$\times$6) surface models constructed by the arrangements of the 3Ga-H (2$\times$2) structure (blue tiles) and Ga$_{ad}$ (2$\times$2) structure (red tiles). Brown, blue and white atoms correspond to Ga, N and H, respectively.}
\end{figure}

More specifically, by using the above approach where DFT calculations for (2$\times$2) lateral cells are combined with thermodynamics, Kusaba \textit{et al.}\ \cite{kusaba2017thermodynamic} identified the GaN(0001) growth surface in MOVPE. 
The most thermodynamically favorable surface is either the Ga adatom surface (Ga$_{ad}$ hereafter), in which a single Ga adatom is adsorbed on the so-called T4 site (see Fig.~\ref{fig:baselines}), or the hydrogen-covered surface (3Ga-H hereafter), in which 3 H atoms are adsorbed on the top-layer Ga atoms (Fig.~\ref{fig:baselines}), depending on the balance of Ga and H$_2$ partial pressures during growth. 
Both structures satisfy the electron counting rule (ECR) \cite{pashley1989electron}: 
All electrons in cation dangling bonds that are energetically unfavorable are transferred to the newly emerged Ga-Ga or Ga-H bonds. 
The satisfaction of ECR on Ga$_{ad}$ and 3Ga-H corresponds to their particular coverages of Ga and H, i.e., $\theta_{\rm Ga}$=0.25 and $\theta_{\rm H}$=0.75. 
If the assumption of the (2$\times$2) periodicity is removed, a mixed adsorption state of Ga and H should appear in a certain growth condition range. 
Hence, we consider that the growth surface is a mixture of Ga-adatom-adsorbed and H-adsorbed areas with those coverages. 
For the modeling of relatively wide areas, we use (6$\times$6) lateral cells (dimensions of 2 nm $\times$ 2 nm). 
However, the Ga- or H-adsorbed atomic configurations in a (6$\times$6) cell with those coverages exceed millions in number (see below), so exploration using conventional DFT calculations is unfeasible. 
In this Letter, we challenge this formidable task by introducing Bayesian optimization in the sequential DFT calculations.

Minimization of a target quantity (energy in the present case) in multidimensional space (atomic coordinates here) is a long-standing issue.
Iterative minimization \cite{payne1992iterative}, such as the conjugate gradient occasionally combined with the simulated annealing technique \cite{kirkpatrick1983optimization}, has been a typical methodology. 
Evolutionary algorithms are also used for several applications, including crystal-structure search \cite{lonie2011xtalopt,zakaryan2017stable,kvashnin2019novel}. 
However, those approaches require a huge number of iterations, thus being inappropriate for the present task. 
Recently, an optimization technique based on Bayesian statistics (Bayesian optimization) \cite{snoek2012practical}, which is possibly capable of predicting the minimum solution from insufficient data, has attracted attention. 
This technique has indeed been applied to material exploration \cite{seko2015prediction,ju2017designing,hou2019machine}, and its validity is partly evidenced. 
Here, we use Bayesian optimization combined with (6$\times$6) lateral-cell DFT calculations in the sequential search of nanometer-scale structures of the GaN growth surface. 
We reach stable structures inaccessible ever through $\sim$\!$10^2$ optimization trials among $\sim$\!$10^5$ candidates.

In this research, we explored the atomic configurations of Ga and H adatoms on a GaN(0001)-(6$\times$6) surface unit cell. 
The surface composition corresponding to Ga$_{ad}$/3Ga-H = 1/2 is considered as an example of intermediate compositions for creating a high-resolution surface phase diagram. 
Pure Ga$_{ad}$ ($\theta_{\rm Ga}$=0.25) and \mbox{3Ga-H} ($\theta_{\rm H}$=0.75) structures have 9 Ga adatoms and 27 H adatoms per (6$\times$6) area, respectively. 
Thus, a structure to be explored has 18 H adatoms and 3 Ga adatoms because of the mixing ratio (Ga$_{ad}$/3Ga-H) of 1/2. 
Choosing these pure Ga-adsorbed and H-adsorbed systems as reference systems, the stability of the structure to be explored can be expressed as the mixing enthalpy:
\begin{eqnarray}
E_{mix} = E[\rm{18H3Ga}] - \frac{2}{3}E[\rm{27H}] - \frac{1}{3}E[\rm{9Ga}]
\label{eq:one},
\end{eqnarray}
where $E[\rm{18H3Ga}]$ is the total energy from the DFT calculations of the GaN(0001)-(6$\times$6) surface slab model with a surface structure to be explored and $E[\rm{27H}]$ and $E[\rm{9Ga}]$ are those of (6$\times$6) surface slab models with pure 3Ga-H and Ga$_{ad}$ surface structures, respectively. 
The final target of this study is to discover a nontrivial (6$\times$6) structure with a lower mixing enthalpy than simple (6$\times$6) structures obtained by patchwork of the two distinct (2$\times$2) structures.

The total energies of the surface slab models were calculated based on real-space density functional theory using the RSDFT package \cite{iwata2010massively,hasegawa2014performance}.
This implementation is suitable for highly parallel computing because fast Fourier transform is unnecessary.
Thus, it has an advantage, especially when applied to large-scale systems. 
The exchange and correlation energies were treated by the Perdew-Burke-Ernzerhof exchange correlation functional using norm-conserving pseudopotentials \cite{perdew1996generalized,troullier1991efficient}.
Ga $3d$ electrons were treated as core electrons. 
The mesh spacing was taken as 0.15 \AA, which corresponds to the cutoff energy of 128 Ry on the conventional plane-wave basis. 
A GaN(0001)-(6$\times$6) surface slab model comprises a vacuum layer of more than 20 \AA~and a GaN film (three bilayers for Bayesian exploration and five bilayers for confirmatory recalculation) with adatoms, in which the bottom surface N dangling bonds are terminated by pseudohydrogens with a charge of 0.75$e$ to mimic a semi-infinite GaN substrate \cite{shiraishi1990new}. 
In surface relaxation, a bottom GaN bilayer with pseudohydrogens was fixed, and the convergence criterion for the forces of \mbox{5$\times10^{-4}$} Hartree/a.u. (atomic unit) was used. 
Visualization of electron density was prepared using the VESTA package \cite{momma2008vesta,momma2011vesta}.

We start with baseline models in which the (6$\times$6) structures are constructed as patchwork of the two distinct structural motifs, the (2$\times$2) 3Ga-H and Ga$_{ad}$ structures. 
Candidate patchwork structures are already dozens in number ($_{9}C_{3}=84$). Among them, we have chosen from {\it our intuition} the following four structures which represent the candidates in terms of the distribution of Ga adtaoms.
The (6$\times$6) structures chosen are shown in Fig.~\ref{fig:baselines}, and the calculated mixing enthalpies are shown in Table~\ref{tab:enthalpies}. 
In structures (a) and (b), the three Ga adatoms (red circles) are located at the vertices of small and large equilateral triangles, respectively. 
Structure (b), in which the Ga adatoms are uniformly distributed in whole lateral plane, is almost energetically comparable to structure (a), in which the Ga adatoms are more localized. 
The Ga adatoms in structures (c) and (d) are on the linear line. 
(Note that structures (b) and (c) are identical to each other in the periodic boundary condition.) 
Structure (d), in which the Ga adatoms are more localized than structures (c), is the most stable, inferring that the localization of Ga adtaoms is energetically favorable.

\begin{figure}
\includegraphics[width=8cm]{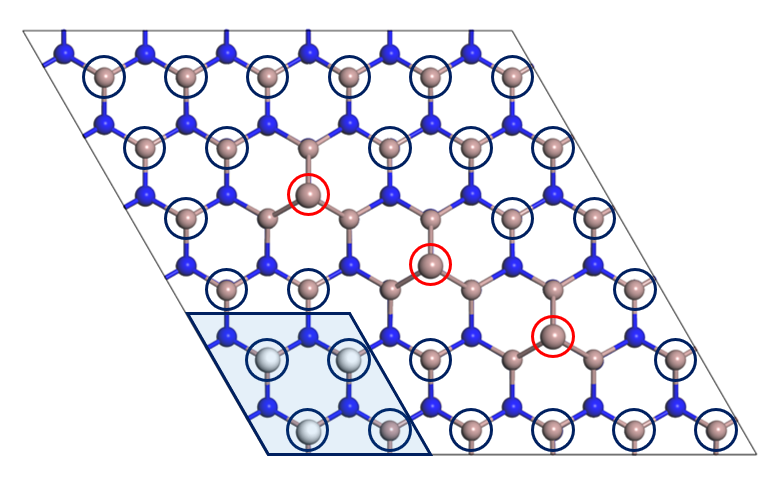}
\caption{\label{fig:class} Top view of a class of (6$\times$6) surface structures to be explored by Bayesian optimization. Red circles indicate Ga adsorption sites and blue circles indicate the candidates for 18 H adsorption sites.}
\end{figure}

We are now in a position to explore the stable (6$\times$6) structures more completely by abandoning the (2$\times$2) structural motifs. Considering the importance of the localization of Ga adtaoms obtained above, we consider a class of surface structures shown in Fig.~\ref{fig:class}, where Ga adatoms are localized in the principal crystallographic direction. 
Keeping the energetically favorable coverages of H atoms and Ga adatoms, i.e., $\theta_{\rm Ga}$=0.25 and $\theta_{\rm H}$=0.75, there are 27 candidate adsorption sites for H, shown as blue circles in Fig.~\ref{fig:class}, among which 18 sites are actually adsorbed. 
We do not consider symmetry constraints in this study since symmetric structures, occasionally assumed from the intuition, are not necessarily most stable. 
Consequently, we encounter with a huge number of candidate structures, i.e., $_{27}C_{18}=4,686,825$ structures.

\begin{table}
\caption{\label{tab:enthalpies} Calculated mixing enthalpies of (6$\times$6) GaN growth surfaces. (a), (b), (c) and (d): The enthalpies of the patchwork models being constructed of the 3Ga-H (2$\times$2) and the Ga$_{ad}$ (2$\times$2) motifs. (\#130): The most stable structure discovered by the present Bayesian optimization. Values in parentheses show recalculated values using the slab models of five GaN bilayers.}
\begin{ruledtabular}
\begin{tabular}{ccc}
Structure name & Preparation & Mixing enthalpy (eV)\\
\hline
a     & Arrangement   & 0.89~~(0.83) \\
b,c     & Arrangement   & 0.83~~(0.84) \\
d     & Arrangement   & 0.54~~(0.54) \\
\#130 & Bayesian Opt. & 0.16~~(0.13) \\
\end{tabular}
\end{ruledtabular}
\end{table}

This challenging exploration is performed by sequential machine learning and adaptive experimental design in the framework of Bayesian optimization. 
The machine learning model takes the surface structure as input and associates it with its total energy as output. 
The optimization is performed with this machine learning model that is updated by Bayesian inference.
Bayesian model allows us to obtain the predicted energy value as well as its uncertainty.
Hence, the optimization generally balances sampling in domain that have not yet been explored and promising domain that have already been explored and are likely to yield good evaluation values, leading to an adaptive sampling strategy to minimize the number of function evaluations. 
In this study, we used an open-source Bayesian optimization library, COMBO \cite{ueno2016combo}, in which the selection of the structure for the next trial is implemented by Thompson sampling \cite{chapelle2011empirical}. 
In exploring the stable structures, we introduce a (2$\times$2) 3Ga-H structural motif (a blue tile in Fig.~\ref{fig:class}) in the (6$\times$6) structure. 
This does not cause loss of generality since this (2$\times$2) motif is expected to be ubiquitous in real surfaces.
On the other hand, this procedure reduces the number, $_{27}C_{18}$, of candidate structures to $_{23}C_{15}=490,314$. 
Each of these candidate surface structures is represented by a 23-dimensional vector in which each row corresponding to each adsorption site is filled with 1 when H is adsorbed and with 0 otherwise. 
This vector is the input to the machine learning model. 
In the COMBO implementation, the input vectors are transformed into feature vectors by a random feature map \cite{rahimi2007random} and input to a Bayesian linear regression model:
\begin{eqnarray}
E_{mix} = \bm{w}^\top \phi(\bm{x}) + \epsilon
\label{eq:two},
\end{eqnarray}
where $\bm{x} \in \mathbb{R}^{23}$ is an input, $\phi \colon \mathbb{R}^{23} \to \mathbb{R}^l$ is a feature map, $\bm{w} \in \mathbb{R}^{l}$ is a weight vector and $\epsilon$ is Gaussian noise. 
The dimension $l$ of the feature vector $\phi(\bm{x})$ was set to 5000. 
This mapping is defined so that, in the limit of $l\to\infty$, the Bayesian linear regression model converges to a Gaussian process \cite{ueno2016combo}. 
The hyperparameters were updated for each trial based on maximization of the type-II likelihood \cite{rasmussen2003gaussian}.

\begin{figure}
\includegraphics[width=8.5cm]{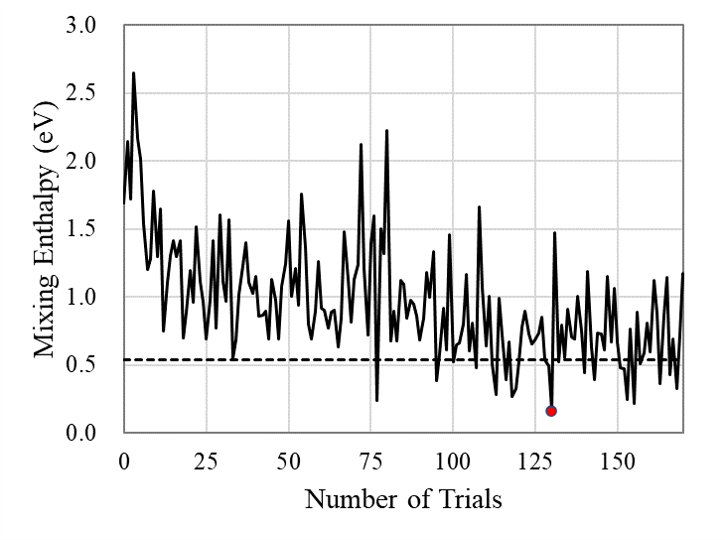}
\caption{\label{fig:trials} History of the exploration of the surface structure by Bayesian optimization. The objective is to minimize the mixing enthalpy. The black dashed line corresponds to the minimum value of the baseline models, and the red marker corresponds to the most stable structure \#130 in the present exploration.}
\end{figure}

Fig.~\ref{fig:trials} shows the results of Bayesian optimization of the H adsorption sites which lower the mixing enthalpy. 
In the early stage of the exploration, when the amount of the training data is insufficient, the mixing enthalpy is so high that it exceeds 2.5 eV. 
Around the trial number \#25, the mixing enthalpy decreases drastically, but a structure more stable than the baseline model (d) is not found yet. 
After that, the trend slowly decreases, indicating that sequential learning of data indeed progresses, although there are sometimes large oscillations. 
After the trial number around \#125, the trend appears to become saturated, so the exploration was terminated at trial number \#170. 
In this exploration, 23 surface structures that are more stable than the baseline model have been discovered: Validity of Bayesian optimization in the present task is evidenced.

\begin{figure}
\includegraphics[width=8.5cm]{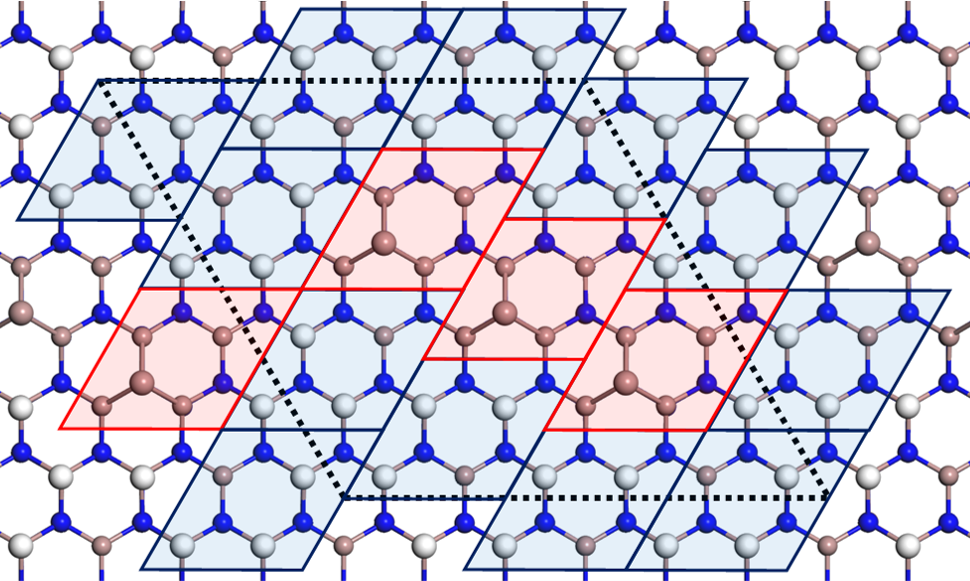}
\caption{\label{fig:structure130} Top view of the stable (6$\times$6) surface structure discovered by the present Bayesian optimization (structure name \#130). The black dashed tile indicates a (6$\times$6) area.}
\end{figure}

The most stable structure discovered in this exploration is structure \#130. 
The top view of the structure is shown in Fig.~\ref{fig:structure130} and its mixing enthalpy value is shown in Table~\ref{tab:enthalpies}.
The structure shows that H adatoms are distributed to avoid the vicinity of Ga adatoms and that the way H adatoms are distributed is asymmetric, resulting in a very complicated adsorption structure. 
One may tend to enumerate symmetrical structures as candidates and explore the most stable structures with limited degrees of freedom. 
This may be related to an ansatz that higher symmetry is favored in nature. 
However, such ansatz is not guaranteed in some phenomena such as crystal growth. 
Hence, the ability to discover complicated surface structure which lacks any symmetry, demonstrated in Fig.~\ref{fig:structure130}, is one of the biggest advantages of using machine learning approaches.

\begin{figure}
\includegraphics[width=6cm]{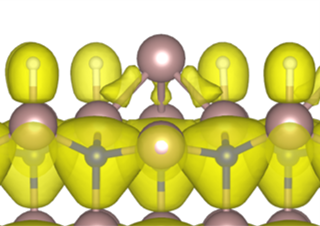}
\caption{\label{fig:density} Calculated electron density near a Ga adatom and H adatoms in the discovered structure \#130. Brown, dark blue and white atoms correspond to Ga, N and H, respectively.}
\end{figure}

Next, with the ECR in mind, let us carefully examine the structure \#130. 
According to the ECR, when one Ga adatom or three H adatoms are adsorbed per (2$\times$2) area in GaN(0001), one of the four topmost Ga atoms has an unoccupied dangling bond, and three of them each form a single bond with the adatom (see Fig.~\ref{fig:density}), resulting in stabilization without excess or deficiency of electrons in the bond network.
In our Bayesian optimization, the structure search is not restricted to patchwork of those (2$\times$2) structural motifs.
However, the resulting structure \#130 clearly shows the dense arrangement of the motifs [red tiles (Ga$_{ad}$) and blue tiles (3Ga-H) in Fig.~\ref{fig:structure130}] with the (3Ga-H) motifs surrounding the (Ga$_{ad}$) motifs.
This finding indicates that the ECR is satisfied locally in the structure \#130 and that further long-range electron transfer is unlikely to occur. 
Fig.~\ref{fig:structure130} shows another interesting feature: The internal configurations, i.e., the positions of the three H atoms, in the (3Ga-H) motifs are different to each other. 
This suggests that the internal configuration is adjusted to reduce the interaction energy among the (2$\times$2) motifs while the ECR is still satisfied.
We argue that the fully local satisfaction of the ECR and the energy reduction due to the internal rearrangement in the motifs, which we have found through the Bayesian optimization, are two essential ingredients to stabilize nanometer-scale GaN surfaces.

In summary, we have identified the stable structure of nanometer-scale GaN (0001) surfaces with Ga- and H-adsorbates, which are fundamental basis for crystal growth modeling, by large-scale density-functional calculations combined with machine learning Bayesian optimization technique. 
We have succeeded to reach a single stable structure with satisfactorily low mixing enthalpy by 130 trials based on Bayesian ptimization among $_{23}C_{15}=490,314$ candidate structures. 
We have found that the obtained structure lacks any postulated high symmetry previously introduced by human intuition, satisfies electron counting rule locally, and shows new adsorbate-rearrangement, leading to the lower mixing enthalpy. 
The present scheme of the Bayesian optimization combined with the first-principle calculations paves a way toward the identification of the surface structures with larger scale and more complex adsorbate-arrangements, and then determination of surface phase diagrams.

\begin{acknowledgments}
This work was partially supported by JSPS KAKENHI (Grant Number JP20K15181) and by MEXT as “Program for Promoting Researches on the Supercomputer Fugaku” (Quantum-Theory-Based Multiscale Simulations toward Development of Next-Generation Energy-Saving Semiconductor Devices, JPMXP1020200205).
Computational resources of the supercomputer system ITO provided by Research Institute for Information Technology, Kyushu University (Project ID: hp200122) are used.
\end{acknowledgments}



%

%

\end{document}